\documentclass{ws-p8-50x6-00}

\begin{document}

\title{Cosmological Evolution of the Universe Neutral Gas Mass
Measured by Quasar Absorption Systems}

\author{Celine Peroux, Richard G. McMahon, Mike Irwin}

\address{Institute of Astronomy, Madingley Road, Cambridge CB3 0HA, UK\\ 
e-mail: celine@ast.cam.ac.uk}

\author{and Lisa J. Storrie-Lombardi}

\address{SIRTF Science Center, California Institute of Technology, MS
100-22, Pasadena CA 91125, USA}  

\maketitle

\abstracts{The cosmological evolution of neutral hydrogen is an
efficient way of tracing structure formation with redshift. It
indicates the rate of evolution of gas into stars and hence the gas
consumption and rate star formation history of the Universe. In
measuring HI, quasar absorbers have proven to be an ideal tool and we
use observations from a recent survey for high-redshift quasar
absorption systems together with data gathered from the literature to
measure the cosmological comoving mass density of neutral gas. This
paper assumes $\Omega_{M}$=0.3, $\Omega_{\Lambda}$=0.7 and
$h$=0.65.}

\section{Quasar Absorption Systems}
Quasar absorbers are systems along the line-of-sight between an
observer and a quasar, and thus are observed in absorption in the
spectra of the background source. They are classified according to
their HI column density: Damped Lyman-$\alpha$ systems (hereafter DLA)
have $N(HI) > 10^{20.3}$ atom cm$^{-2}$, Lyman Limit Systems
(hereafter LLS) have $N(HI) > 10^{17.2}$ atom cm$^{-2}$ and the
lyman-$\alpha$ forest is composed of systems with smaller column
densities.  The work presented here is based on observations from a
recent high-$z$ survey (Peroux et al. 2001) together with a
compilation of absorption systems from the literature
(Storrie-Lombardi and Wolfe 2000).
\begin{figure}[t]
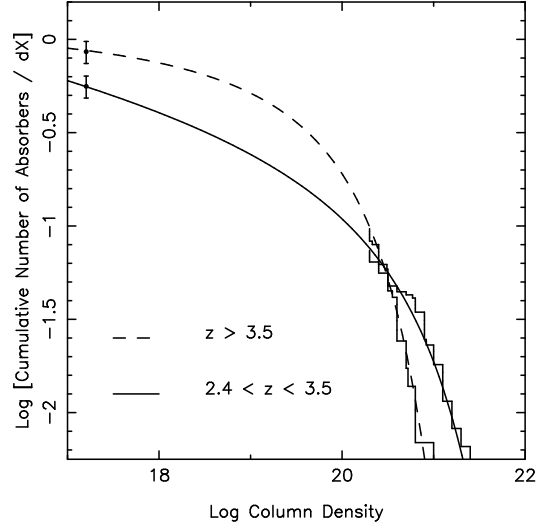

\epsfxsize=16.5pc 
\figurebox{20pc}{15pc}{peroux_f1.eps} 
\caption{\label{f1}Cumulative number of absorbers per unit distance
interval, $dX$, for two different $z$ ranges. The ``steps'' correspond to
the observed number of DLAs and the data points at $\log N(HI) = 17.2$
are the expected number of LLS derived from the observed number of LLS
per unit $z$. $\Gamma$-functions are used for the fits to the data.}
\end{figure}
This included 82 quasars with $z > 4$ and resulted in a sufficiently
large sample of quasar absorption systems to enable the study of their
properties with redshift ($z$). Fig 1 shows the cumulative number of
quasar absorbers per unit $dX(z)$, for two $z$ ranges, absorbers with
$z < 3.5$ and $z > 3.5$, where $dX$ is the distance interval:
$X(z)=\int_{0}^{z}(1+z)^2 \left[ (1+z)^2 ( 1 + z \Omega_M) - z ( 2 + z
) \Omega_{\Lambda} \right]^{-1/2} dz$.  The column density of LLS
cannot be derived directly from their equivalent widths as they
correspond to the non-linear part of the curve of
growth. Nevertheless, the expected number of LLS can be calculated by
fitting $N_o(1+z)^{\gamma}$ to the observed number of systems per unit
$z$ and integrating over the $z$-path surveyed: $LLS_{expected}=
\sum_{i=1}^{n} \int_{z_{min}}^{z_{max}}N_o(1+z)^{\gamma}dz$. The
observations presented in Fig 1 show that there are fewer systems at
the highest column density for $z > 3.5$ compared to the lower
redshift range, which indicates that we are probing the epoch of
assembly of high column density systems from smaller column density
units.

\section{Cosmological Evolution of Neutral Hydrogen}
\begin{figure}[t]
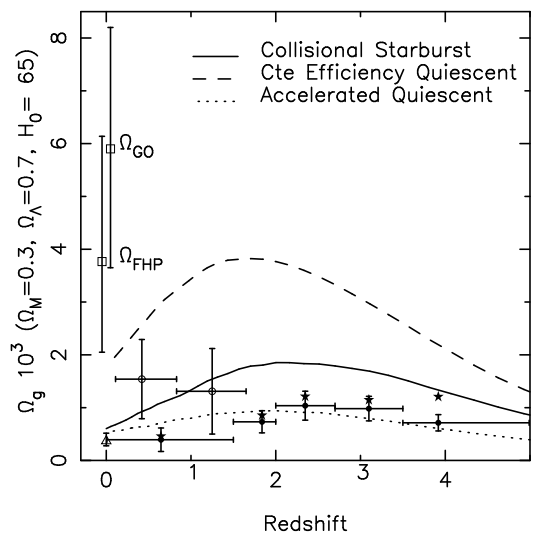

\epsfxsize=16.5pc 
\figurebox{20pc}{20pc}{peroux_f2.eps} 
\caption{\label{f2}Cosmological evolution of HI. The circles show the
neutral gas contained in DLAs (filled circles are this work, open
circles are from Rao and Turnshek 2000 who used HST spectra of quasars
with know MgII systems to look for low-$z$ DLAs. Their results are
surprisingly high but might be biased by small number statistics or the
effect of gravitational lensing.) Vertical error bars correspond to
1$\sigma$ uncertainties and the horizontal error bars indicate bin
sizes. The stars are the total HI including a correction for the
neutral gas contained in systems with column densities below the DLA
formal definition. The triangle at $z=0$ is the local HI mass measured
by Rao $\&$ Briggs 1993. The squares, $\Omega_{FHP}$ and $\Omega_{GO}$
(Fukugita, Hogan $\&$ Peebles 1998 and Gnedin $\&$ Ostriker 1992
respectively) are $\Omega_{baryons}$ in local
galaxies. Semi-analytical models from Somerville, Primack $\&$ Faber
2000 are overplotted.}
\end{figure}

The neutral gas mass is calculated by integrating the quasar
absorbers column density distribution, $f(N,z)$: $\Omega_{HI}(z) =
\frac{H_o \mu m_H}{c \rho_{crit}} \int_{N_{min}}^{\infty} N f(N,z)
dN$.  We used a $\Gamma$-function to fit the data (Fig 1) in order to
take into account the HI included below the formal DLA
definition. Indeed, DLAs were previously thought to contain the
majority of the total HI but our new study shows that 14$\%$ at $z <
3.5$ and up to 40$\%$ at $z > 3.5$ of the neutral gas is actually
contained in smaller column density systems ($10^{17.2} < N(HI) <
10^{20.3}$). The total HI decreases with decreasing z thus matching
the pattern expected from gas depletion due to star formation.
Details on the method used and more discussion on the results
presented here will be available in a subsequent paper.


\begin{thebibliography}{99}
\bibitem{} Fukugita, M., Hogan, C. and Peebles, P. 1998, ApJ, 503,
518.
\bibitem{} Gnedin, N. and Ostriker, J., 1992, ApJ, 400, 1.
\bibitem{} Peroux, C., Storrie-Lombardi, L., McMahon, R.,
Irwin, M. and Hook, I., 2001, AJ, in press, astro-ph/0101179.
\bibitem{} Rao, S. and Briggs, F. 1993, ApJ, 419, 515.
\bibitem{} Rao, S. and Turnshek, D. 2000, ApJS, 130, 1.
\bibitem{} Somerville, R., Primack, J. and Faber, S. 2000, MNRAS,
in press, astro-ph/0006364.
\bibitem{} Storrie-Lombardi, L. and Wolfe, A. 2000, ApJ, 543, 552.
\end{thebibliography}
\end{document}